\documentclass[graybox, natbib, nosecnum, twocolumn]{svmult}
\bibpunct{(}{)}{;}{a}{}{,} 

\usepackage{mathptmx}       
\usepackage{helvet}         
\usepackage{courier}        
\usepackage{type1cm}        

\usepackage{makeidx}         
\usepackage{graphicx}        
\usepackage{multicol}        
\usepackage[bottom]{footmisc}
\usepackage[normalem]{ulem}	
\usepackage{hyperref}  
\usepackage{soul}   

\usepackage{amsmath,amssymb,mathrsfs}
\usepackage[normalem]{ulem}

\def\bc{\mbox{blockchain}}
\usepackage{caption, subcaption}

\newcommand{\bft}{\textsc{BFT}}
\newcommand{\POW}{\textsc{PoW}}
\newcommand{\POS}{\textsc{PoS}}
\newcommand{\POA}{\textsc{PoA}}
\newcommand{\POC}{\textsc{PoC}}
\newcommand{\resilientdb}{\textsc{ResilientDB}}
\newcommand{\pbft}{\textsc{Pbft}}
\newcommand{\zyzzyva}{\textsc{Zyzzyva}}
\newcommand{\SBFT}{\textsc{Sbft}}
\newcommand{\hotstuff}{\textsc{Hotstuff}}
\newcommand{\PoE}{\textsc{PoE}}
\newcommand{\RBFT}{\textsc{Rbft}}
\newcommand{\multibft}{\textsc{Rcc}}
\newcommand{\GeoBFT}{\textsc{GeoBft}}
\newcommand{\expodb}{\textsc{ResilientDB}}

\newcommand{\Client}{\mathcal{C}}
\newcommand{\Replica}[1]{\mathcal{#1}}
\newcommand{\Primary}{\mathcal{P}}
\newcommand{\m}{{\tt m}}
\newcommand{\MName}[1]{\textsc{#1}}

\begin{document}

\title*{Blockchain Transaction Processing}
\author{Suyash Gupta, Mohammad Sadoghi}
\institute{Suyash Gupta, Mohammad Sadoghi 
\at University of California Davis, 
\email{(sugupta,msadoghi)@ucdavis.edu}}
%
%
\maketitle
\section{Synonyms}
\begin{itemize}
\item Blockchain Data Management
\item Blockchain Consensus
\item Cryptocurrency
\end{itemize}

\section{Definitions}
A blockchain is an append-only linked-list of blocks,
which is maintained at each participating node. 
Each block records a set of transactions and their 
associated metadata.
Blockchain transactions act on the identical ledger
data stored at each node.
Blockchain was first perceived by 
Satoshi Nakamoto~\citep{bitcoin} as a peer-to-peer 
digital-commodity (also known as crypto-currency) exchange system.
Blockchains received traction due to their inherent property of 
immutability---once a block is accepted, it cannot be reverted.

\section{Overview}
In 2008, Satoshi Nakamoto~\citep{bitcoin} introduced the
design of an unanticipated technology that revolutionized
the research across the distributed systems community.
Nakamoto presented the design of a peer-to-peer  
digital-commodity exchange system, which although employed by several 
participants, prevents the use of a centralized design.
Nakamoto envisioned a system where the participants exchange commodities 
among themselves in a democratic, decentralized and transparent manner
while upholding their right to privacy.
Nakamoto visualized this digital-commodity as a {\em monetary} token that could 
be used by participants to provide or receive services.
This led to the birth of {\em Bitcoin}---a cryptocurrency---and introduction 
of a new design paradigm {\em Blockchain}.

\begin{figure}[t]
\centering
\includegraphics[width=0.9\columnwidth]{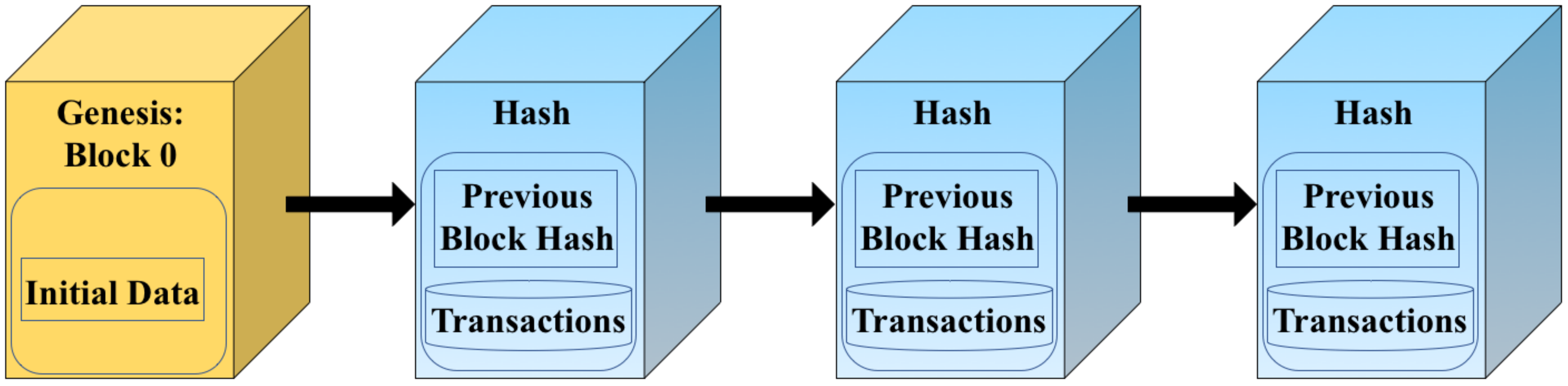}
\caption{\small Basic Blockchain Representations}
\label{fig:blockchain}
\end{figure}

A blockchain in its simplest form is an {\em append-only} linked-list of blocks.
Each block in this chain is linked to the previous block in the chain~\cite{middleware-tutorial,vldb-tutorial}. 
Blockchains are often termed as immutable as modifying an existing block requires 
modifying all the previous blocks in the chain.
Each block includes a set of transactions and the associated meta-data.
Figure~\ref{fig:blockchain} presents a schematic representation of a \bc{}.
Blockchain systems guarantee decentralization as the full-copy of the chain is maintained 
by several participants%
\footnote{In sharded blockchain systems, no shard may have complete copy but data is 
still securely replicated.}.
Moreover, a block is only accepted into the chain after all the participants 
have reached consensus on the {\em order} and {\em contents} of the block.
In specific, admittance of a block to the chain implies that the 
transactions iin the block have been executed and verified.
Hence, blockchain helps in achieving key properties such as democracy and transparency.

A blockchain system can be described as a collection of layers. 
At the {\em application layer}, there are clients, which send their 
transactions to a set of severs to process. 
The communication among the clients and servers take place at the networking layer. 
Servers participate at the {\em ordering layer} to assign a unique order 
to each incoming client transaction in a {\em Byzantine Fault-Tolerant} 
(henceforth referred to as \bft{}) manner. 
Following a successful ordering, the client transaction is processed at the 
{\em execution layer} and persisted in the immutable ledger at the {\em storage layer}.
Clients and servers also employ necessary cryptographic constructs to securely 
exchange messages among each other.

The preceding discussion allows us to summarize that 
a \bc{} system aims at providing a {\em safe} and {\em resilient} 
storage for transactions. 
In the succeeding sections, we will discuss these concepts in detail 
and will illustrate the mechanisms pertaining to blockchain transaction processing. 
We will also study key principles required to order and validate these client transactions 
and provide analysis of some existing \bc{} applications.

\section{Key Research Findings}
Each blockchain system can be visualized as a secure representation of 
a traditional {\em database} system~\citep{nawab2018}.
Similar to a database system, each blockchain application also receives
transactions from multiple clients. 
In its vanilla form, a \bc{} transaction  
is a collection of {\em read or write} 
operations.
Clients issue these transactions to the 
servers for processing and exchange of digital-commodities.

In general, each server in a blockchain application stores a 
full-copy of the chain. 
Hence, without any loss of generality, we can claim that servers of a 
blockchain system are {\em replicas} of each other. 
In specific, a blockchain application lays down a replicated design where 
each replica participates in ordering and executing the incoming client transaction.

Prior works have shown that it is possible to make a replicated system handle 
failures~\citep{paxos,distsystem-maarten}. 
In any replicated system, replicas participate in a fault-tolerant {\em consensus} 
protocol to decide the order to execute a client transaction. 
Blockchain applications also adhere to this philosophy. 
They employ a \bft{} consensus protocol to achieve consensus under byzantine failures.
But, {\em why do BFT protocols need to handle byzantine failures?}
As a blockchain system promotes democracy, it permits display of adversarial behavior by malicious replicas during consensus. 
To tackle such malicious activities, each blockchain application relies on the design and 
properties dictated by a \bft{} consensus protocol.

%
\begin{figure*}[t]
\begin{subfigure}{0.45\textwidth}
	\includegraphics[width=\textwidth]{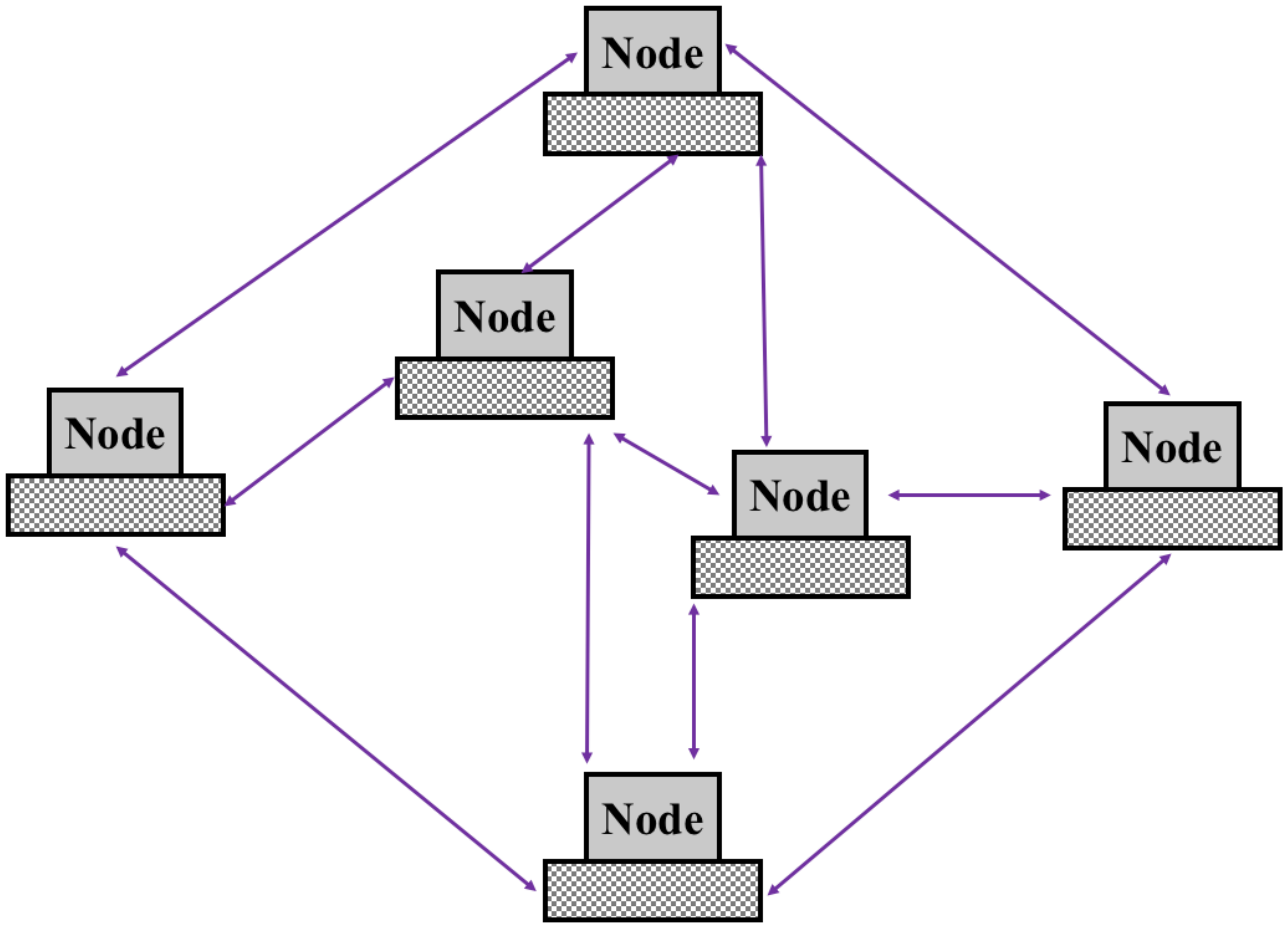}
	\caption{Public Blockchain.}
	\label{fig:publicbc}
\end{subfigure}%
\hfill
\begin{subfigure}{0.45\textwidth}
	\includegraphics[width=\textwidth]{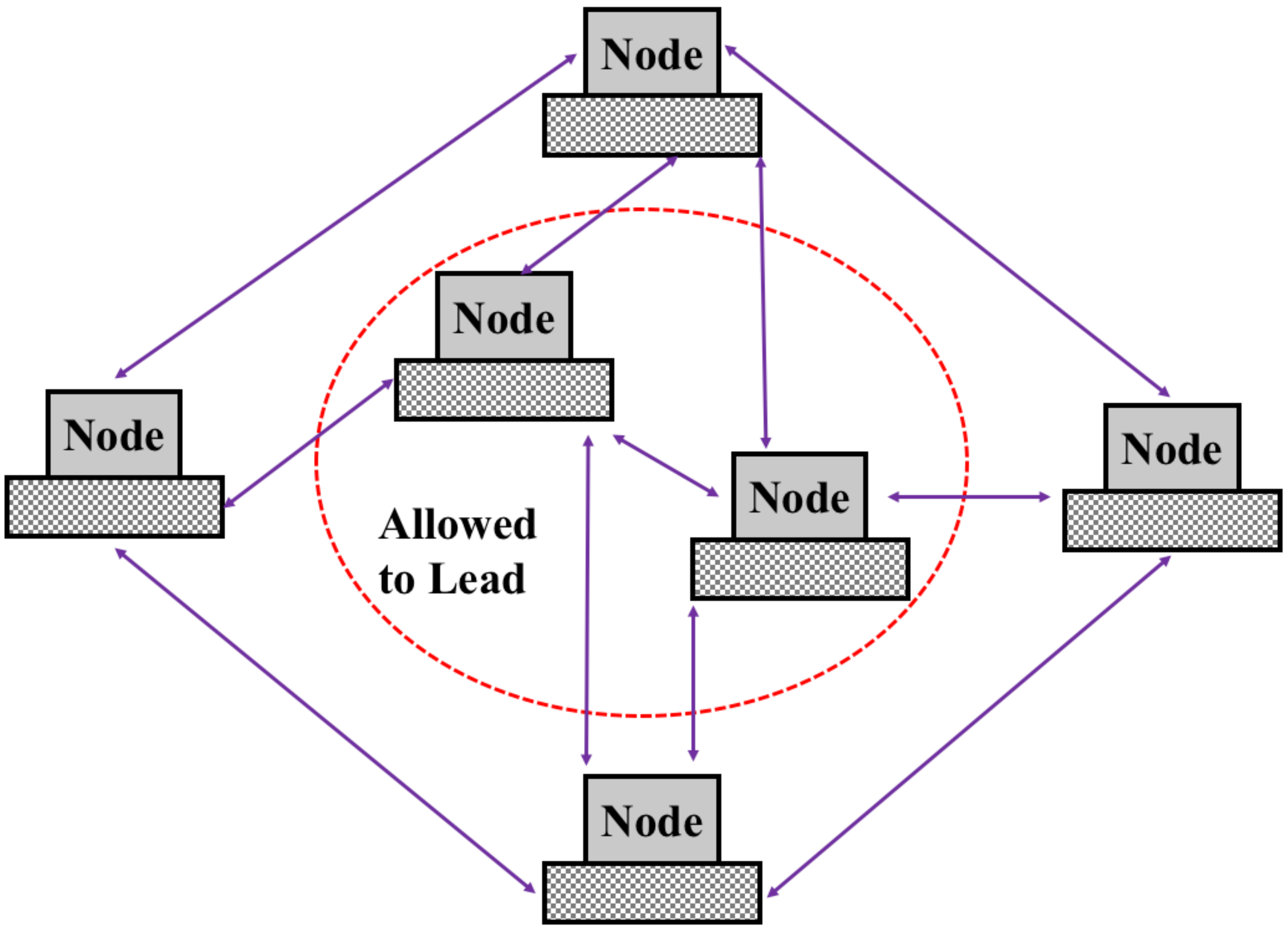}\hfill
	\caption{Hybrid Blockchain.}
	\label{fig:hybridbc}
\end{subfigure}%
\hfill
\begin{subfigure}{0.45\textwidth}
	\includegraphics[width=\textwidth]{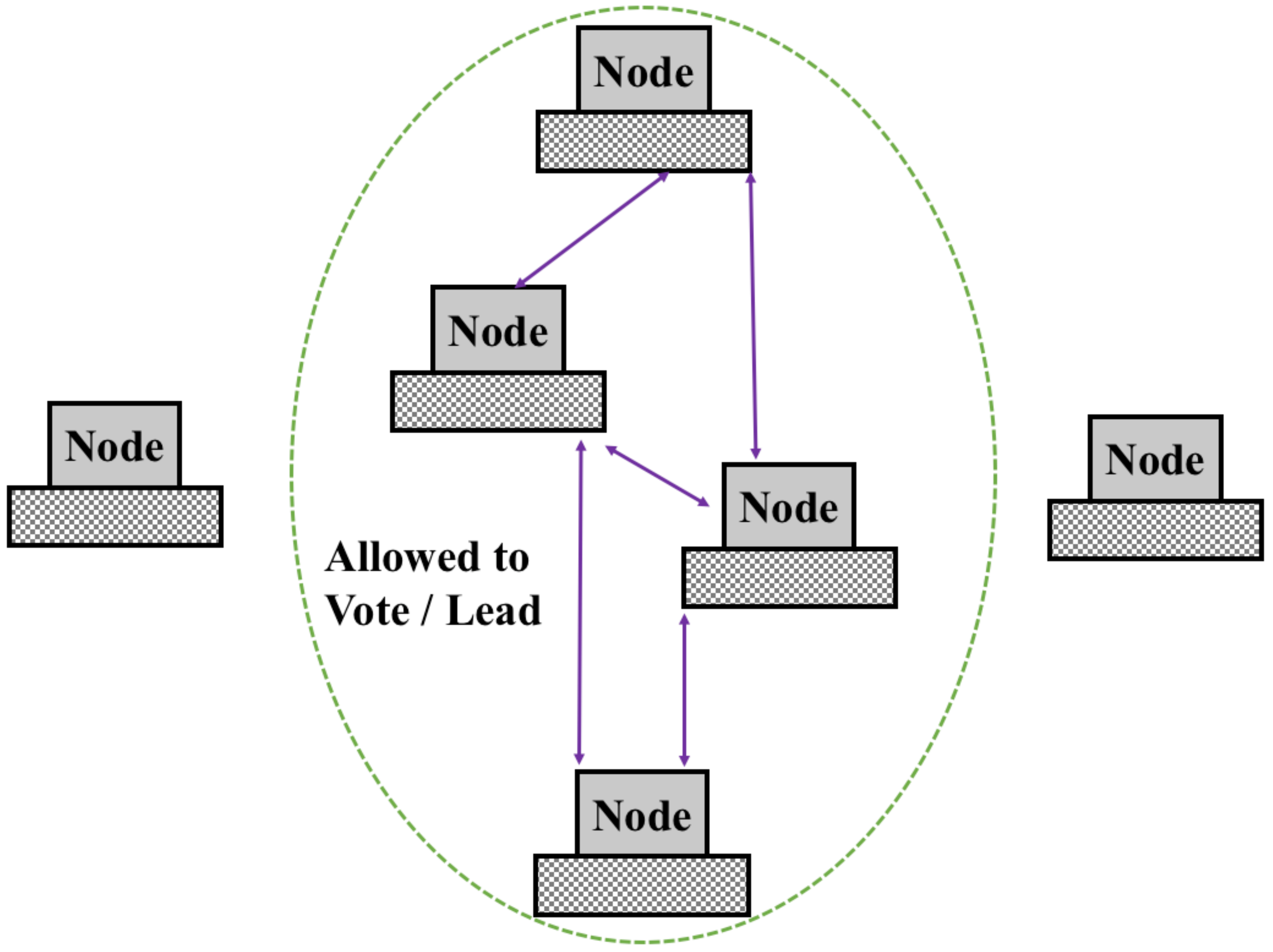}
	\caption{Permissioned Blockchain.}
	\label{fig:permissionedbc}
\end{subfigure}%
\hfill
\begin{subfigure}{0.45\textwidth}
	\includegraphics[width=\textwidth]{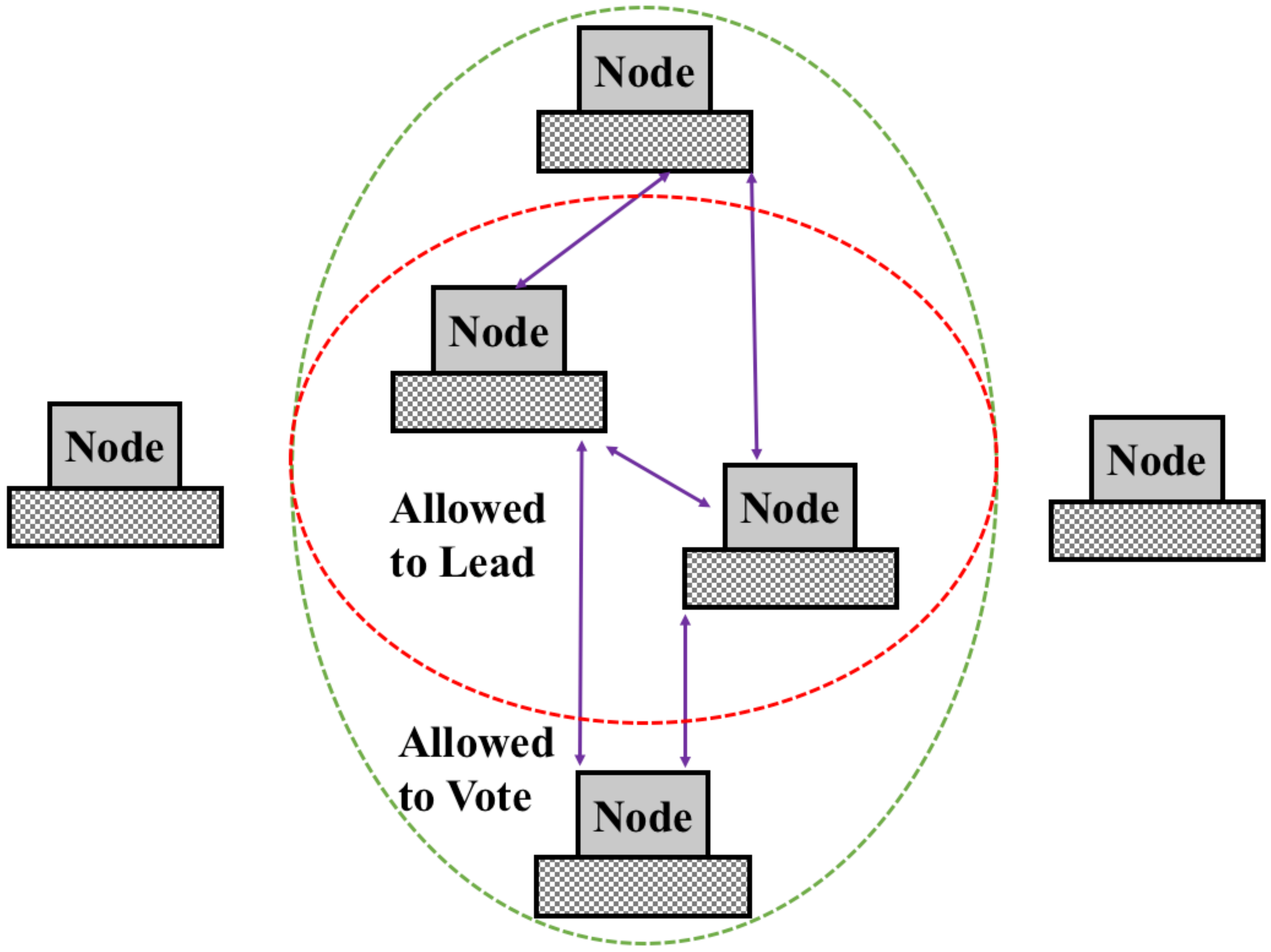}\hfill
	\caption{Private Blockchain.}
	\label{fig:privatebc}
\end{subfigure}%
\caption{Topologies for Blockchain Systems.}
\label{fig:classify-blockchain}
\end{figure*}
\section{Blockchain Topologies}
A key parameter that renders the design of a \bc{} system 
is its underlying application.
On the basis of permissions available to a participating node, 
a blockchain application can be categorized \bc{} as permissioned, permissionless  
or hybrid~\citep{blockchaintech,blockchain-consensus}. 
Although the \bc{} community agrees on the characteristics of a permissionless 
or {\em public} \bc{} infrastructure, 
there is a lack of concise definitions to explain other models.

On the basis of topology, we categorize a \bc{} systems under four heads: 
public, private, permissioned and hybrid.
Figure~\ref{fig:classify-blockchain} presents a 
pictorial representation of the different categories.
In these figures, nodes that lack any connections are disallowed from participating 
in the management of the blockchain.
Further, we use circles to demarcate 
different zones of operation; 
certain nodes are allowed to lead the consensus (or create the next block) while
some nodes are allowed to participate in the consensus protocol.

{\em Public Blockchain} systems, such as 
Bitcoin~\citep{bitcoin} and Ethereum~\citep{ether},
allow any node to participate in the consensus process 
and propose the next valid block for the chain.
Hence, a public or permissionless blockchain system upholds its democratic nature 
by providing each node with equal probability%
\footnote{%
The equal probability of creating a block is only guaranteed when all the nodes have 
exactly same amount of resources, and  
each node is working independently.}
of creating the next block to be added to the chain.

{\em Private Blockchain} systems run at the other extreme end of 
the spectrum. 
These blockchain systems permit only a specific set of nodes to be part of 
the consensus protocol and restrict the creation of next block to an even 
smaller subset of nodes.
Private blockchain designs are attractive to large multi-sector companies and banks, 
which may chose to allow some of their customers to participate in the consensus protocol, 
while restricting creation of next block to its employees.

{\em Hybrid Blockchain} systems attain a middle ground 
between the two extremes. 
Although these systems allow any node to be part of the consensus 
protocol, they restrict the task of proposing and creating the next block to a 
designated subset of replicas.
For instance,  Ripple~\citep{ripple}---a cryptocurrency---supports 
a variant of the hybrid model. 
In Ripple, only some public institutions have the permissions to 
select the transactions that will be part of the next block.

Amidst all these topologies, {\em permissioned blockchain} systems have 
successfully created a niche space for their design~\citep{hyperledger-fabric,resilientdb}.
Permissioned blockchain applications allow any node participate in the consensus protocol 
but require the identities of all participants to be known a priori. 
Although participants loose their privacy, permissioned blockchain applications 
provide each participant equal opportunity to propose the next block.
Notice that permissioned blockchain applications place no other special restrictions on 
the behavior of a participant.
Hyperledger Fabric~\citep{hyperledger-fabric}, Libra coin~\citep{libra} and \resilientdb~\citep{resilientdb} 
are some of the state-of-the-art permissioned blockchain applications and fabrics.

\section{Blockchain Transactional Flow}
The initial block of any blockchain is  
termed as the {\em genesis block}~\citep{infopropagation}.
Genesis block is a special block that is numbered zero, and 
is hard-coded in every \bc{} application.
Each other block links to some previously 
existing block.
Hence, a \bc{} grows by appending new blocks 
to the existing chain.

A transaction in a \bc{} system is identical to any 
distributed or OLTP transaction~\citep{tpcc} that acts 
on some data. 
Traditional \bc{} applications (such as
Bitcoin) consist of transactions that 
represent an exchange of money between two entities 
(or users).
Each valid transaction is recorded in a block, which can 
can contain multiple transactions, for efficiency.
Immutability is achieved by leveraging strong
cryptographic properties such as hashing~\citep{cryptobook}.

\begin{figure*}
\centering
\includegraphics[width=0.8\textwidth]{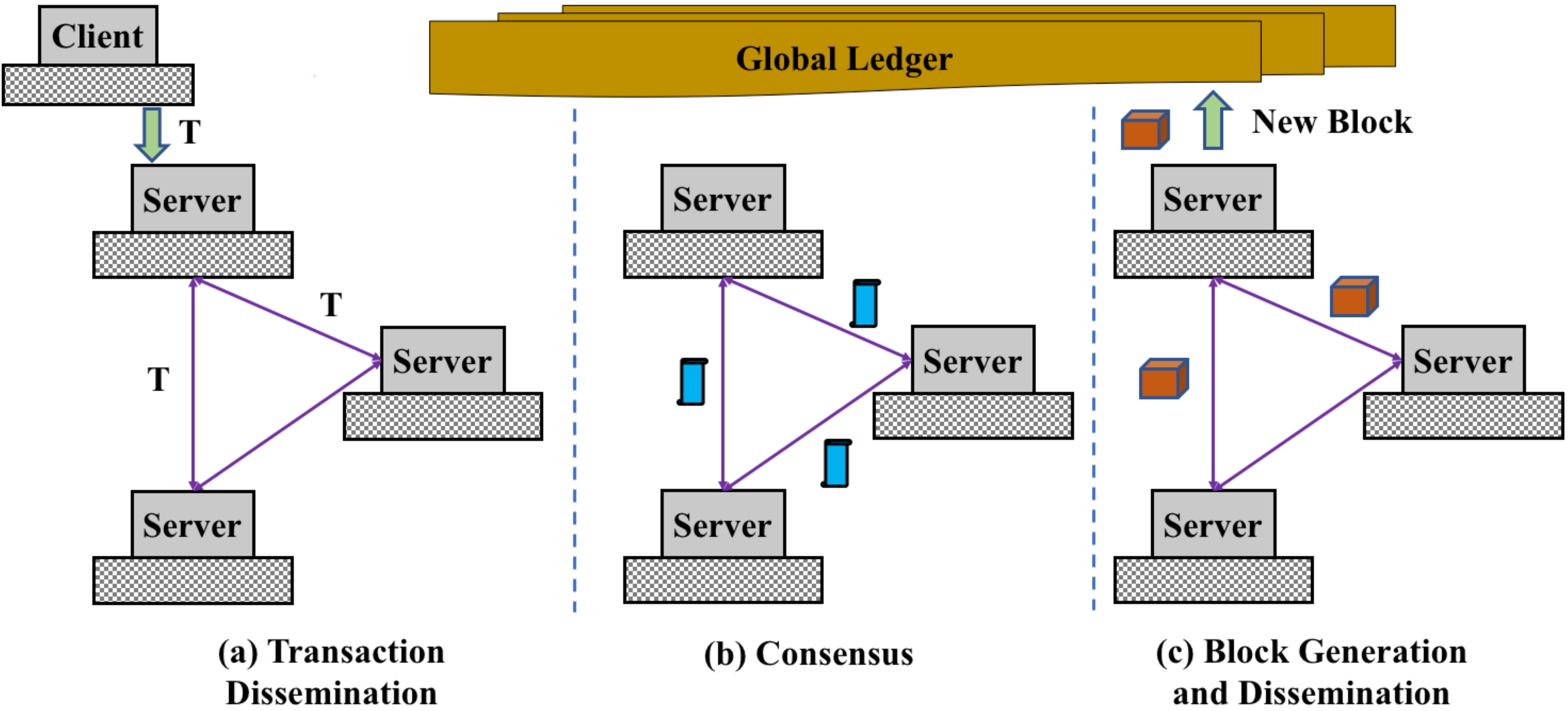}
\caption{\small {\bf Blockchain Flow:} Three main phases in any blockchain 
application are represented. 
(a) Client sends a transaction to one of the server, which it 
disseminates to all the other servers. 
(b) Servers run the underlying consensus protocol, to determine 
the block creator. 
(c) New block is created, and transmitted to each node, which also 
implies adding to global ledger.}
\label{fig:blockflow}
\end{figure*}

Figure~\ref{fig:blockflow} illustrates the 
three main phases 
required by any \bc{} application to create a new block.
The client transmits a transactional request to one 
of the participants. 
This participating node multicasts the client request to all other nodes. 
We term this phase as {\em Transaction Dissemination}. 
Once, all the nodes have a copy of client request, 
they initiate a {\em consensus} protocol. 
The choice of underlying consensus protocol affects 
the time complexity and resource consumption. 
The winner of the consensus phase proposes the 
next block and transmits it to all other nodes. 
This transmission process is equivalent to adding 
an entry (block) to the global distributed ledger.

\section{Blockchain Consensus}
At the core of any blockchain application is a \bft{} consensus protocol which states that 
{\em given a client transaction, the aim of this consensus protocol is to ensure all the 
non-faulty replicas assign the same order to this transaction.} 
Depending on the underlying topology, we can broadly categorize consensus protocols 
into two categories: permissionless consensus protocols and permissioned consensus protocols.

Achieving fault-tolerant distributed consensus is an age-old problem. 
Commit protocols such as {\em Two-Phase Commit}~\citep{2pc}, {\em Three-Phase Commit}~\citep{3pc} 
and {\em EasyCommit}~\citep{easyc,easyc-extend} help in reaching agreement among the 
participants in a {\em partitioned} distributed databases~\citep{quecc,qstore,tbook}.
However, commit protocols can only handle node failures and are {\em unsafe} under 
message delay or loss.

Paxos~\citep{paxos} and Viewstamped Replication~\citep{viewstamp} 
allow a distributed system of replicas to achieve consensus  
in the presence of {\em crash-faults}.
In a system of $n$ replicas, a system employing Paxos for consensus 
can handle up to $n$ failures where $n \ge 2f+1$.
Notice that these $f$ failures need not be simple replica crashes but can also 
take form of message losses and delays.
However, crash-fault tolerant protocols such as Paxos and Viewstamped Replication 
cannot handle any malicious behavior.

A byzantine-fault tolerant protocol aims at reaching consensus in a system of $n$ 
replicas where at most $f$ replicas can act as byzantine and $n \ge 3f+1$.
Traditional \bft{} protocols promote a {\em primary-backup} model where one 
replica is designated as the primary and other replicas act as backups. 
It is the task of the primary to initiate consensus among all the backups.
Notice that all the above discussed protocols, such as Two-Phase Commit, Paxos and so on, 
follow the primary-backup model.
The key reason primary-backup model is preferred is because of its simplicity and its 
ability to {\em blame the primary} for an unsuccessful consensus.
 
Recent \bc{} applications
present several new protocols for achieving 
consensus: Proof-of-Work~\citep{pow, bitcoin}, 
Proof-of-Stake~\citep{ppcoin} and
Proof-of-Authority~\citep{parity}. 
Prior works have shown that these consensus protocols 
provide similar guarantees as traditional \bft{} protocols~\citep{pbftequiv}.
Hence, in the rest of this section, we illustrate some of 
the state-of-the-art blockchain protocols for 
both permissioned and permissionless systems.

\subsection{Permissioned Consensus}

A decade prior to the inception of the 
first \bc{} application, 
the problem of achieving fault-tolerant 
distributed consensus problem had already excited 
practitioners and researchers~\citep{paxos,viewstamp,pbft}. 
Distributed systems research community agreed that 
a byzantine-fault tolerant system can only be considered 
correct if it is both {\em safe} and {\em live}. 
A replicated system is called as {\em safe} if all its replicas 
are consistent, that is, have the same state.
A replicated system is termed as {\em live} if its replicas 
are able to make progress, that is, process incoming client requests.

A majority of existing \bft{} protocols guarantee safety 
under {\em asynchronous} environment, that is, messages can 
get loss, delayed or duplicated, and 
up to $f$ replicas may act byzantine.
Further, any \bft{} protocol employs cryptographic 
constructs to prevent malicious replicas from impersonating 
non-faulty replicas. 
As clients send their transactions to other replicas,  
so each client uses {\em digital signatures} to sign its message~\citep{hac,cryptobook}.
For all other messages, depending on the algorithmic steps, 
the system can employ either {\em asymmetric-key} digital signatures 
or less-expensive {\em symmetric-key} message authentication codes~\citep{cryptobook}.
Hence, we assume {\em authenticated communication}: 
malicious replicas can impersonate each other, 
but no replica can impersonate a non-faulty replica.
Further, replicas will accept only those messages which are 
well-formed, that is, have valid message authentication 
codes or digital signatures (as applicable).

{\bf PBFT.} 
{\em Practical Byzantine Fault Tolerance}~\citep{pbft} 
if often considered as the first protocol to present a 
practical design for achieving byzantine fault-tolerance in a 
distributed system. 
\pbft{} follows the primary-backup model where the primary replica
initiates the consensus among all the replicas. 
It is the responsibility of the primary to ensure all the 
backup replicas 
successfully order every incoming client transaction otherwise 
it risks replacement.
If the primary is non-malicious and the network is reliable,
\pbft{} guarantees consensus in {\em three} phases.

\pbft{} protocol starts when a client $\Client$ wants a 
transaction to be executed and sends a {\em request} 
$\m$ to the primary replica $\Primary$. 
The primary $\Primary$ checks if the client signature is valid 
and if this is the case, it creates a $\MName{Pre-prepare}$ 
message and sends that message to all the backups.
This $\MName{Pre-prepare}$ message includes a sequence 
number (an integer) and a {\em hash} of the client request.
The sequence number $k$ states the order to execute the transaction 
while the hash acts as a digest, which can be used in future 
communications as an alias for the client request%
\footnote{
Client requests are often of the order of several kilobytes and 
sending an hash instead optimizes the communication.
}.

When a replica $\Replica{R}$ receives a $\MName{Pre-prepare}$ message 
from the primary $\Primary$, it performs the following 
checks:
(i) verifies the client signature on $\m$, 
(ii) checks if $\Primary$ is the primary, and 
(iii) ensures the sequence number $k$ has not already been used.
If the $\MName{Pre-prepare}$ message passes all the checks, 
$\Replica{R}$ agrees to support primary's order for this request 
and sends a $\MName{Prepare}$ message to all the replicas.
When a replica $\Replica{R}$ receives $\MName{Prepare}$ messages 
from $2f$ replicas in support of the request $\m$ sent by 
$\Primary$, then $\Replica{R}$ marks the request as {\em prepared}. 
This information gives $\Replica{R}$ an assurance that a majority 
of non-faulty replicas are also agreeing to order this request at 
sequence $k$.
Next, $\Replica{R}$ acknowledges the prepared request by sending 
a $\MName{Commit}$ message to all the replicas.
When a replica $\Replica{R}$ receives $\MName{Commit}$ messages 
from $2f+1$ replicas, then $\Replica{R}$ achieves a unique guarantee 
on the order of $m$, that is, a majority of non-faulty replicas 
have also prepared $m$.
This allows replica $\Replica{R}$ to go ahead and execute the 
request $\m$ as the $k$-th request. 
Finally, $\Replica{R}$ sends the result of executing $m$ 
as a response to the client $\Client$.

The client $\Client$ needs $f+1$ matching responses from 
distinct replicas, to mark its request $\m$ as complete. 
It is possible that the client may not receive sufficient 
number of matching responses. 
To handle such cases, the client initiates a timer prior 
to sending its request. 
In specific, each client waits on a timer for receiving $f+1$ 
identical responses. 
If the client {\em timeouts} while waiting for 
$f+1$ responses, then it forwards its request $\m$
to all the replicas. 
When a {\em backup} replica $\Replica{R}$ 
receives a client request $\m$, 
it forwards that request to the primary $\Primary$ 
and starts its timer. 
If $\Primary$ fails to send a $\MName{Pre-prepare}$ 
message corresponding to $\m$, then $\Replica{R}$ 
concludes that $\Primary$ is byzantine and initiates 
primary replacement.
Existing literature terms this primary replacement process 
as {\em view-change} because each primary represents a 
{\em view} of the system.
The view-change protocol only starts when at least $f+1$ 
replicas are ready to replace the primary. 
This condition is necessary as up to $f$ replicas can be byzantine 
and may even request replacement of a non-faulty primary. 
Hence, when at least $f+1$ replicas request replacement, 
remaining replicas assume that there is at least one 
non-faulty replica which has been affected.

For a successful view-change to take place, a new primary has 
to be selected. 
\pbft{} follows a simple principle: if the replica 
with index $i$ is the current primary, then replica with index 
$j$ will be the next primary, where $j = (i+1) \bmod n$. 
But, how does a replica concludes that it is time for it to act 
as the {\em new} primary. 
When any replica $\Replica{R}$ receives $\MName{View-Change}$ 
messages from $2f+1$ distinct replicas that want to elect it as the 
primary, then it initiates the process of switching to next view.
Notice that the process of switching to next view requires ensuring 
all the replicas have the common state. 
Thus, the new primary also needs to provide this information as part 
of the $\MName{New-View}$ message.

{\bf Zyzzyva.}
It is evident from \pbft's design that it requires three phases of 
communication of which two necessitate quadratic communication 
complexity.
Hence, there is a need for optimized protocols, which can achieve 
the same goals with much lesser communication overheads.
\zyzzyva~\citep{zyzzyva} presents a {\em twin-path} protocol that 
achieves consensus in a single linear phase if there are no failures.
All the replicas in the \zyzzyva{} start in the {fast-path} and 
switch to the {\em slow-path} under failures.
Note that a recent work has illustrated that \zyzzyva{} is 
unsafe under failures~\citep{zyzzyva-unsafe}.

In \zyzzyva{}, when a non-primary replica $\Replica{R}$ receives a 
$\MName{Pre-prepare}$ message from the primary $\Primary$, it 
assumes that the primary is non-faulty and agrees to execute 
this request. 
Such an execution is termed as speculative as the replica $\Replica{R}$
is unaware of the state at other replicas. 
In specific, a byzantine primary could have equivocated and sent 
different replicas distinct client requests.
Once the replica $\Replica{R}$ executes the request, it sends the 
reply to client $\Client$. 
The client $\Client$ marks the request complete if it receives 
matching identical responses from at least $3f+1$ replicas.

A keen reader can easily notice that the onus is on the client to 
ensure system is safe.
Further, when $n = 3f+1$, then the client has to wait for responses from 
all the replicas.
Due to these restrictions, \zyzzyva's fast-path works only if there 
are no failures.
In \zyzzyva{}, the client waits on a timer while expecting $3f+1$ 
responses. 
If the client timeouts prior to receiving responses, then it 
initiates the {\em slow-path}. 
In the slow-path, client has to summarize the state it received from 
different replicas and needs to decide whether primary replacement 
needs to be initiated or a simple recovery protocol
is sufficient to ensure system remains live. 
Clearly, the slow-path is no longer linear and 
requires multiple phases.
Moreover, if the client is malicious, then the replicas could be 
momentarily unsafe until there is a good client.
Another key challenge with twin-path protocols is finding the 
optimal timeout value.
Prior works have shown that finding a timeout value can be hard and 
\zyzzyva{} faces severe reduction is throughput under 
failures~\citep{upright,aadvark,poe}.

{\bf \SBFT.} 
The key aim behind the design of \SBFT~\citep{sbft} is to make a 
consensus protocol that can guarantee safe consensus with 
linear message complexity in periods of no failures.
In fact, like \zyzzyva{}, \SBFT{} is also a twin-path protocol.
\SBFT{} employs {\em threshold signatures} to achieve linear 
communication complexity.

Threshold signatures are based on asymmetric cryptography. In specific, each replica holds a distinct {\em private key}, which it can use to create a signature {\em share}. 
Next, one can produce a valid {\em threshold signature} given 
at least $t$ such signature shares from distinct replicas
(the exact value of $t$ is dependent on the underlying consensus protocol).

At a closer look, it seems like 
\SBFT{} requires more phases than \pbft{}. 
This occurs because \SBFT{} linearizes each phase of \pbft{} through use of threshold signatures.
In \SBFT{}, when a replica $\Replica{R}$ receives a 
$\MName{Pre-prepare}$ message, it agrees to support
from the primary's sequence by generating a threshold share. 
The replica $\Replica{R}$ sends this share to a specific replica 
designated as the {\em collector}.
When a collector receives message from 
at least $3f + 2c + 1$ replicas it generates a threshold 
signatures and sends this signature to all the replicas.
When a replica receives a threshold signature from the collector, 
it executes the request to generate a response, 
creates a threshold share on this response and 
sends these to a specific replica designated as 
the {\em executor}.
The executor waits for $f+1$ identical responses and 
combines them into threshold signature. 
Next, the executor sends this 
signature to all the replicas and clients.

For \SBFT's fast path to work as stated, either there should be {\em no failures} or 
at least $3f + 2c + 1$ replicas should participate in consensus where up to $c > 0$ replicas can crash-fail (no byzantine failures).
Moreover, the primary can act as both collector and executor but \SBFT{} suggests using distinct replicas in fast path.
If the collector {\em timeouts} waiting for threshold shares from 
$3f + c + 1$ replicas, it switches 
to the slow path, which requires two additional linear phases 
to complete consensus.

{\bf HotStuff.}
In any primary-backup \bft{} protocol, if the primary acts malicious, then the protocols employ 
the accompanying view-change algorithm to detect and replace the malicious primary. 
This view-change algorithm leads to a momentary disruption in system throughput until the resumption of service.

\hotstuff{}~\citep{hotstuff} proposes eliminating the dependence of a \bft{} consensus protocol from 
one primary by replacing primary at the end of every consensus.
Although this rotating leader design escapes the cost of a view-change protocol, 
it enforces an implicit {\em sequential} paradigm. 
Each primary needs to wait for its turn before it can propose a new request.

In \hotstuff{}, in round $i$, the replica with identifier 
$i \bmod n$ acts as the primary and proposes a request to 
all the replicas.  
Each replica on receiving this request, creates a threshold 
share and sends to the replica $\Replica{R}$ 
with identifier $(i+1) \bmod n$.
If $\Replica{R}$ receives threshold shares from $2f+1$ replicas, 
then it combines them into a threshold signature 
and initiates the consensus for round $i+1$ by 
broadcasting its proposal along with the computed threshold signature.
Notice that replicas have not executed the request and replied to the client. 
\hotstuff's aim is to linearize the consensus proposed by 
\pbft{} protocol, which it does by 
splitting each phase of \pbft{} into two using threshold signatures.
To reduce the communication, it chains the phases.
Hence, a replica executes the request for the $i$-th round once it receives a threshold signature
from the primary of $(i+3)$-th round.
Evidently, chaining helps \hotstuff{} to some extent but it does not eliminate its sequential nature.
This sequential nature forces \hotstuff{} to loose out on an opportunity to process messages out-of-order.

{\bf PoE.}
{\em Proof-of-Execution} (henceforth referred 
to as \PoE{}) consensus protocol 
aims at achieving consensus in three 
linear phases without relying on any 
twin-path model~\citep{poe}. 
Further, \PoE{} recognizes that 
no one size fits all systems~\citep{no-one-bft}.
Hence, its design is independent 
of the choice of {\em underlying} cryptographic signature scheme.
This implies that the \PoE{} protocol 
can employ  
both symmetric and asymmetric-cryptographic 
signature schemes depending on the 
application environment.

The design of \PoE{} is built on three 
key insights. 
First, \PoE{} prevents use of any 
twin-path paradigm as switching from 
fast to slow-path requires 
dependence on timeouts, which degrades system performance.
Second, \PoE{} allows replicas to speculatively execute the requests but 
facilitates rollbacks in case of inconsistencies.
Final, \PoE{} allows out-of-order processing, which eliminates any 
bottlenecks associated with sequential consensus protocols.

For the sake of brevity, we will describe \PoE{} built on top of threshold signatures.
In \PoE{}, the client $\Client$ initiates execution by sending its request $\m$ to the primary $\Primary$. To initiate replication and execution of $\m$ as the $k$-th transaction, the primary proposes $\m$ to all replicas by broadcasting a $\MName{Propose}$ message.

After a replica $\Replica{R}$ receives a $\MName{Propose}$ message from $\Primary$, 
it checks whether at least $2f$ other replicas also received the same proposal from $\Primary$. 
To perform this check, each replica agrees to \emph{support} the first $k$-th proposal 
it receives from the primary 
by sending a $\MName{Support}$ message that includes its unique {\em threshold share} to the primary. 
The primary $\Primary$ waits for $2f+1$ threshold shares, and on receiving such shares, 
it combines them into a {\em threshold signature} and broadcasts as a \MName{Certify} message.
When a replica $\Replica{R}$ receives the \MName{Certify} message, 
it \emph{view-commits} to $\m$  as the $k$-th transaction in view $v$. 
After $\Replica{R}$ view-commits to $\m$, $\Replica{R}$ schedules speculative execution of $\m$. 
Consequently, $\m$ will be executed by $\Replica{R}$ after all preceding transactions are executed. 
After execution, $\Replica{R}$ informs the client of the order of execution and 
of any execution result. 
A client considers its transaction successfully executed after it receives identical
response messages from $2f+1$ distinct replicas.

{\bf Aardvark.}
The design philosophy behind Aardvark 
is distinct in comparison to existing 
\bft{} protocols~\citep{aadvark}.
It aims at building a {\em robust} 
\bft{} protocol that can continue performing 
under failures.
Hence, in the failure-free cases, 
Aardvark attains lower throughput than 
a majority of the existing \bft{} protocols.

In Aardvark, 
prior to sending its request to the primary, 
the client signs the request using both 
digital signatures and message authentication 
codes. 
This prevents malicious clients from 
performing a 
{\em denial-of-service} attack 
as it is expensive for 
client to sign each message twice.
Aardvark also employs a
{\em point-to-point} network 
rather than the multicast network 
for exchange of messages among clients 
and replicas. 
The key intuition behind such a choice 
is to disallow a faulty client or replica 
from blocking the complete network.
Aardvark also periodically changes the primary replica. 
Each replica tracks the throughput of the current 
primary and suggests replacing the primary 
when there is a decrease in its throughput.
To perform such tracking, each replica 
sets a timer and measures the rate of 
primary's responses.

{\bf RBFT.}
The key intuition behind the design of \RBFT{}
is to facilitate detection of {\em clever} 
malicious primaries~\citep{rbft}.
\RBFT{} extends Aardvark 
and aims to detect those malicious primaries, 
which cannot be detected by simple timers 
suggested by Aardvark.
 
In Aardvark, a clever primary can avoid 
detection by delaying messages 
just slightly below the timeout threshold.
Such a primary can throttle the system 
throughput without risking eviction.
To tackle this challenge, 
\RBFT{} insists running $f+1$ 
independent instances of the 
Aardvark protocol on each replica. 
One of these instances is designated as 
the {\em master} while other 
instances act as {\em backups}.
Irrespective of the designation of an
instance, all the instances order all the 
requests.
However, only the master instance executes 
the requests.

The key task of the backup instances is 
to monitor the performance of the master 
instance. 
If any backup instance observes a 
degradation of the 
system throughput at the master, 
it broadcasts a message to 
elect a new primary. 
Further, to guarantee at least one of the 
$f+1$ instances is led by a non-faulty 
replica, \RBFT{} requires each instance to be 
led by a distinct replica.
In comparison to both \pbft{} and Aardvark, 
\RBFT{} requires an additional 
phase, which is used to propagate the client 
requests across all the replicas.

{\bf RCC.}
Although \RBFT{} successfully utilizes redundancy to detect clever malicious primaries, 
it also wastes excessive bandwidth by requiring all the instances to order the same set of requests.
{\em Resilient Concurrent Consensus} (henceforth referred to as \multibft{}) paradigm resolves 
this issue by parallelizing the consensus~\citep{discmbft,multibft}.
In specific, \multibft{} runs at each replica {\em multiple instances} of a primary-backup protocol. 

The key challenge with the design of primary-backup protocols is their reliance
on the primary. 
This dependence can severely affect the throughput and scalability of these protocols.
The primary replica not only receives all
client requests but is also responsible for ensuring consensus is 
reached on the order for these requests among all other replicas.
If the primary {\em fails} to ensure consensus, then all remaining replicas 
need to {\em replace} this primary.
This replacement process is necessary as, without it, non-faulty replicas 
may never converge. 
Unfortunately, primary replacement is not cheap, as it requires pausing consensus on all outstanding requests until the 
primary is replaced.

\multibft{} aims at making a \bft{} consensus primary agnostic. 
To achieve such a property, \multibft{} advocates running $z$ parallel instances at each replica.
Further, \multibft{} ensures that each instance is managed by a 
distinct replica.
Using parallelization, \multibft{} ensures that the non-faulty replicas are {\em always accepting and ordering client requests}, this independent of any malicious behavior or attack.

We now present the design of \multibft{} paradigm that parallelizes
the seminal \pbft{} consensus protocol. 
For the sake of explanation, we assume \multibft{} works in {\em rounds}.
Each {\em round} of \multibft{} includes {\em three stages}: {\em parallel consensus}, 
{\em unification}, and {\em execution}.
The notion of a {\em round} helps in generating a common order and 
recovering from instance failures but it does not prevent individual primaries 
from working independently.

Prior to any round, \multibft{} requires each replica to prepare to run $z$ 
instances of \pbft{} protocol in parallel.
A round $r$ begins when the primary of each instance proposes a client request.
Firstly, in the {\em parallel consensus} stage, each instance 
runs \pbft{} on its client request. 
Secondly, in the {\em unification} stage, the replica waits for all its 
$z$ instances to complete replication (reach consensus on their respective requests).
If every instance successfully replicates a request, 
then a common order for execution of these requests is determined.
If one or more instances are {\em unable to replicate} requests, 
then the primaries for those instances must be faulty and recovery is initiated.
Finally, in the {\em execution} stage, each replica 
executes all the client requests in the common order.

{\bf GeoBFT.}
Existing \bft{} protocols do not distinguish between the \emph{local} and \emph{global} communication, 
which is a necessary requirement to enable  geo-scale deployment of a blockchain system.
To resolve this challenge, {\em Geo-Scale Byzantine Fault-Tolerant} 
consensus protocol (henceforth referred to as \GeoBFT{}) 
that uses topological information to group all replicas in a single region into a single cluster~\citep{geobft}. 
Likewise, \GeoBFT{} assigns each client to a single cluster. 
This clustering helps in attaining high throughput and scalability in geo-scale deployments. 
\GeoBFT{} operates in rounds, and in each round, every cluster will be able to propose a single client request for execution. 
Each round consists of the three steps: \emph{local replication}, \emph{global sharing}, and \emph{ordering and execution}, which 
we further detail next.

At the start of each round, each cluster chooses a single transaction of a local client. 
Next, each cluster \emph{locally replicates} its chosen transaction in a Byzantine fault-tolerant manner using \pbft{}. 
At the end of successful local replication, \pbft{} guarantees that each non-faulty replica can prove successful local replication via a \emph{commit certificate}.

Next, each cluster shares the locally-replicated transaction along with its commit certificate with all other clusters.  
To minimize inter-cluster communication, we use a novel \emph{optimistic global sharing protocol}. 
Our optimistic global sharing protocol has a global phase in which clusters exchange locally-replicated transactions, followed by a local phase in which clusters distribute any received transactions locally among all local replicas. 
%
Finally, after receiving all transactions that are locally-replicated in other clusters, 
each replica in each cluster can deterministically \emph{order} all these transactions and proceed with their \emph{execution}.  
After execution, the replicas in each cluster inform only local clients of the outcome of the execution of their transactions 
(e.g., confirm execution or return any execution results).

\subsection{Permissionless Consensus}
Permissionless applications inspired by Nakamoto's 
Bitcoin~\citep{bitcoin}
advocate a public blockchain system where 
any replica can participate in the consensus.
Hence, the identity of a participating replica can be protected.
This design property requires the underlying consensus protocol 
used to order the transactions to expend the resources of a 
participant.
In specific, each participant needs to spend some 
of its resources if it wants to propose the next block.
If such a resource consumption is not enforced, then a 
malicious participant can create multiple pseudonymous identities 
and subvert the system, also known as the Sybil attack~\citep{sybil-attack}.

{\bf Proof-of-Work.}
Bitcoin relies on the {\em Proof-of-Work} (henceforth referred 
to as \POW{}) protocol to achieve consensus among a set of replica. 
\POW{} protocol builds on top of a simple intuition 
``What is mathematically hard to compute but easy to verify?''
Hence, \POW{} protocol requires the computation to be expensive, 
that is, it should deplete some resources of the {\em prover}.

In \POW{} protocol, the participating nodes compete 
among themselves to propose the next block by solving a 
complex puzzle.
In nature, several computationally hard problems exist, such as Diophantine Equation, RSA Factorization, 
One-way Hash Functions, and so on.
Among these hard problems, following the Nakamoto's vision, 
\POW{} protocol is associated with the computation of 
{\em one-way hash functions} such as 
computation of a $256$-bit SHA3 value. 
When a node $N$ successfully computes this hash value, it 
disseminates this solution to all other nodes for verification.
Any node can verify this solution to check $N$'s claim.

The main critic behind \POW's design is that leads to excessive 
wastage of energy. 
Permissionless applications that employ \POW{} consensus have to 
set large {\em targets} to prevent Sybil attacks. 
Further, \POW's design facilitates unfair practices---higher the 
computational capabilities a node has higher are its chances of 
solving the complex puzzle. 
Such a design promotes {\em pooling of resources} where several 
nodes work together to compute the hash.
Moreover, a node has to be given {\em incentives} to participate 
in the \POW{} consensus.
If the incentives are not sufficient, then nodes may decline 
creating the next block, which in turn can either stall the system
or compromise its security.

Another issue with the \POW{} consensus protocol is that it 
can lead to tricky situations where it is hard to determine 
the next block in the chain.
For instance, two nodes $N_1$ and $N_2$ may solve the complex 
puzzle at the same time. 
In such a case, it is possible that one half of the remaining 
participants may receive a solution from $N_1$ before $N_2$ while 
the other half receives solution from $N_2$ before $N_1$. 
To handle this scenario, some form of resolution mechanism is needed, 
which would lead to wastage of resources of either $N_1$ or $N_2$ 
as both of their blocks cannot be appended to the chain.
Notice that any new block added to the chain includes the hash of the 
previous block.


{\bf Proof-of-Stake.} 
In \POW{} protocol, miners have to deplete their computational resources in order to earn 
the right to create the next block. 
Each miner who controls a fraction $s$ of the total computational power, has a probability 
nearly equal to $s$ to create the next block.

{\em Proof-of-Stake} (henceforth referred to as \POS{}) presents a principle that contrasts 
the resource usage philosophy of \POW{}.
In a blockchain system employing \POS{} protocol, a replica possessing a higher stake than the
other replicas gets a chance to create a new block~\citep{pos-bentov}.
In specific, the probability a replica possessing a fraction $s$ of the total stakes in the system 
creates the next block is $s$.
The key security rationale behind \POS{} is that replicas who have some stake 
involved in the system are also {\em well-suited} to ensure its security.

PPCoin or PeerCoin~\citep{ppcoin} is often regarded as the first implementation of \POS{}. 
The key motivation behind PPCoin's design was to implement a crypto-currency that does not 
require participating replicas to spend its resources in performing large computations.
Initial \POS-based design were based on the notion of {\em coinage}. 
In specific, a replica's ability to create the next block is determined on its value of coinage.
Coinage is calculated on the basis of number of days a replica has held some coins or stake.
To prevent Sybil attacks, \POS-based systems require replicas to 
 algorithm requires a node to spend its coinage if it wants to 
propose the next block.

Initial implementations of the \POS{} protocol lacked the fairness 
criterion.
This is evident as the replica with the highest stake gets the 
chance to propose the next block.
Although a high stake replica looses its coinage once it 
creates the next block, it may create the subsequent block 
if its stake is much larger in value than that of the other replicas.

To resolve this issue, a {\em chain-based} variant of 
\POS{} algorithm has been proposed. 
The chain-based \POS{} protocol employs a psuedo-random 
algorithm to select a validator, which then creates 
a new block and adds it to the existing chain of blocks.
The frequency of selecting the validator is set to 
some pre-defined time interval.
Another variant of \POS{} algorithm follows {\em BFT-style} consensus.
In this design, the replicas participate in a 
\bft{} protocol to select the next valid block.
Here, validators are given right to propose the next 
block, at random.
The key difference between these algorithms is 
the synchrony requirement; {\em chain-based} \POS{} algorithms 
are inherently synchronous, while {\em BFT-style} \POS{} 
is partially synchronous.

Another {\em key} challenge for \POS-based designs is an 
attack by {\em rational} stakeholders.
A rational replica would always aim at maximizing its profit, 
an expected behavior in a democracy in correspondence with the Nash equilibrium~\cite{pos-bentov}.
Rational replicas can affect the security of \POS{}, as in at attempt to 
maximize their gains, they may participate in multiple chains.

A rational miner could get blocks from distinct {\em forks} of the blockchain. 
To maximize its returns, a miner would attempt to propose the next block 
for each such fork. 
As miners don't lose any actual resources (like computational energy in \POW{}),
so they are free to propose blocks on different chains.
This could lead to an ever-expanding divergent network.

{\bf Proof-of-Authority.}
A variation of \POS{} algorithm to be employed in hybrid 
blockchain topologies is termed as {\em Proof-of-Authority} 
(henceforth referred as \POA{})~\citep{parity}.
The key idea is to designate a set of nodes as the 
authorities or leaders. 
These authorities are entrusted with the task of creating 
new blocks and validating the transactions.
\POA{} marks a block as part of the \bc{} if it is 
signed by majority of the authorized nodes.
The incentive model in \POA{} highlights that it is in 
the interest of an authority node to maintain its 
reputation. 
In case an authority acts malicious, it can 
loose its status and periodic incentives.
Hence, \POA{} does not select nodes based on 
their claimed stakes.

{\bf Proof-of-Space.}
A consensus algorithm orthogonal to the design proposed by 
\POW{} is {\em proof-of-space} or {\em proof-of-capacity} 
(henceforth referred as \POC{})~\citep{posateniese,postefan}. 

\POC{} expects nodes to provide a proof that they have 
sufficient ``storage'' to solve a computational problem. 
\POC{} algorithm targets computational problems 
such as {\em hard-to-pebble graphs}~\citep{postefan} 
that need large amount of memory storage to solve 
the problem. 
In the \POC{} algorithm, the verifier first expects a 
prover to commit to a labeling of the graph, and 
then it queries the prover for random locations in the 
committed graph.
The key intuition behind this approach is that unless 
the prover has sufficient storage, it would not pass
the verification.
SpaceMint~\citep{spacemint}---a
\POC-based~cryptocurrency---claims that \POC{} based 
approaches are more resource efficient in comparison 
to \POW{} as storage consumes less energy.

\section{Blockchain Systems}
We now briefly look at the design of some of the state-of-the-art 
blockchain applications and fabrics.
The key aim of this section is to illustrate the different design practices 
adopted by existing blockchain systems.

{\bf Bitcoin}~\citep{bitcoin} is regarded as the 
first ever \bc{} application. 
It is a cryptographically secure digital currency designed with the 
aim of disrupting the traditional institutionalized 
monetary exchange.
Bitcoin acts as the token of transfer between two parties 
undergoing a monetary transaction. 
The underlying \bc{} system is a network of nodes 
(also known as {\em miners}) that take a set of client 
transactions and validate the same by 
demonstrating a {\em proof-of-work}, that is 
generating a block.
The process of generating the next block is non-trivial 
and requires large computational resources. 
Hence, the miners are given incentives (such as Bitcoins) 
for dedicating their resources and generating the block. 
Each miner maintains locally an updated copy of the 
complete \bc{} and the associated ledgers for every 
Bitcoin user.

To ensure Bitcoin system remains fair towards all the machines, 
the difficulty of {\em proof-of-work} challenge is periodically 
increased.
Prior works have illustrated that Bitcoin is vulnerable to $51\%$ attack, which 
can lead to {\em double spending}~\citep{bitrate}. 
The intensity of such attacks increases when multiple forks 
of the longest chain are created. 
To avoid these attacks, Bitcoin developers 
suggest the clients to wait for their block 
to be confirmed before they mark the Bitcoins as 
transferred.
This wait ensures that the specific 
block is a little
deep (nearly six blocks) in the longest 
chain~\citep{bitrate}.
Bitcoin critics also argue that its {\em proof-of-work} 
consumes huge energy~\footnote{
As per some claims one Bitcoin transaction 
consumes power equivalent to that required by 
$1.5$ American homes per day.
}
and may not be a viable solution for future. 


{\bf Ethereum}~\citep{ether} is \bc{} framework that permits 
users to create their own applications ({\em smart-contracts}) on top of the 
Ethereum Virtual Machine (EVM). 
Ethereum utilizes the notion of smart contracts to 
facilitate development of new operations. 
It also supports a digital cryptocurrency, {\em ether}, 
which is used to incentivize the developers to create 
correct applications. 
One of the key advantage of Ethereum is that it  
supports a Turing complete language to generate new 
applications on top of EVM. 
At the time of writing this article, Ethereum employs 
a variant of \POW{} protocol to achieve consensus among 
its miners.
Ethereum makes its miners
solve challenges that were not only computational intensive, 
but also memory intensive. 
This design prevented existence of miners who utilized 
specially designed hardware for compute intensive applications.

In future, Ethereum Foundation aims to switch to a variant 
of \POS{} protocol to reach consensus among its replicas.
The modified protocol is referred to as 
{\em Casper}~\citep{casper}.
Casper introduces the notion of {\em finality}, 
that is, it ensures that one chain becomes 
permanent in time.
It also introduces the notion of {\em accountability},
which penalizes any validator that attempts the 
{\em nothing-at-stake} attack.
The penalty leveraged on such a validator is 
equivalent to negating all his stakes.

{\bf Parity}~\citep{parity} is an application designed on 
top of Ethereum. 
It provides an interface for its users to interact with the 
Ethereum \bc{}. 
Parity allows its \bc{} community 
to use either {\em Proof-of-Work} and {\em Proof-of-Authority} 
to reach consensus in their applications. 
Hence, if some users select \POA{} consensus, then 
Parity provides mechanisms for  
setting up the {\em authority} nodes.

{\bf Ripple}~\citep{ripple} is considered as third largest 
cryptocurrency after Bitcoin and Ethereum in terms of market cap.
It employs a consensus algorithm which is a simple variant of 
existing traditional \bft{} algorithms. 
Ripple requires number of failures $f$ to be 
bounded as follows: $\leq (n-1)/5 + 1$, where $n$ represents 
the total number of nodes.
Ripple's consensus algorithm introduces the notion of 
a {\em Unified Node List} (UNL), which is a subset of the network. 
Each server communicates with the nodes in its UNL for 
reaching a consensus.
The servers exchange the set of transactions they received 
from the clients and propose those transactions to their 
respective UNL for vote.
If a transaction receives $80\%$ of the votes, it is marked 
permanent. 
Notice that if the generated UNL 
groups are a clique then forks of the longest chain could 
co-exist. 
Hence, UNLs are created in a manner that they share some 
set of nodes.
Another noteworthy observation about Ripple protocol 
is that each client 
needs to select a set of validators or unique nodes that they 
trust. 
These validators utilize the ripple consensus algorithm to 
verify the transactions.

{\bf Hyperledger}~\citep{hyperledger} is a suite of 
resources aimed at modeling industry standard \bc{} 
applications. 
It provides a series of Application Programming Interfaces
(APIs) for developers to create their own 
non-public \bc{} applications. 
Hyperledger provides implementations of \bc{} 
systems that uses RBFT and other variants of the 
PBFT consensus algorithm.
It also facilitates use and development of smart contracts. 
It is important to understand that the design philosophy of 
Hyperledger leans towards \bc{} applications that 
require existence of non-public networks, and 
so, they do not need a compute intensive consensus.

\begin{figure*}[t]
\centering
\includegraphics[width=0.7\linewidth]{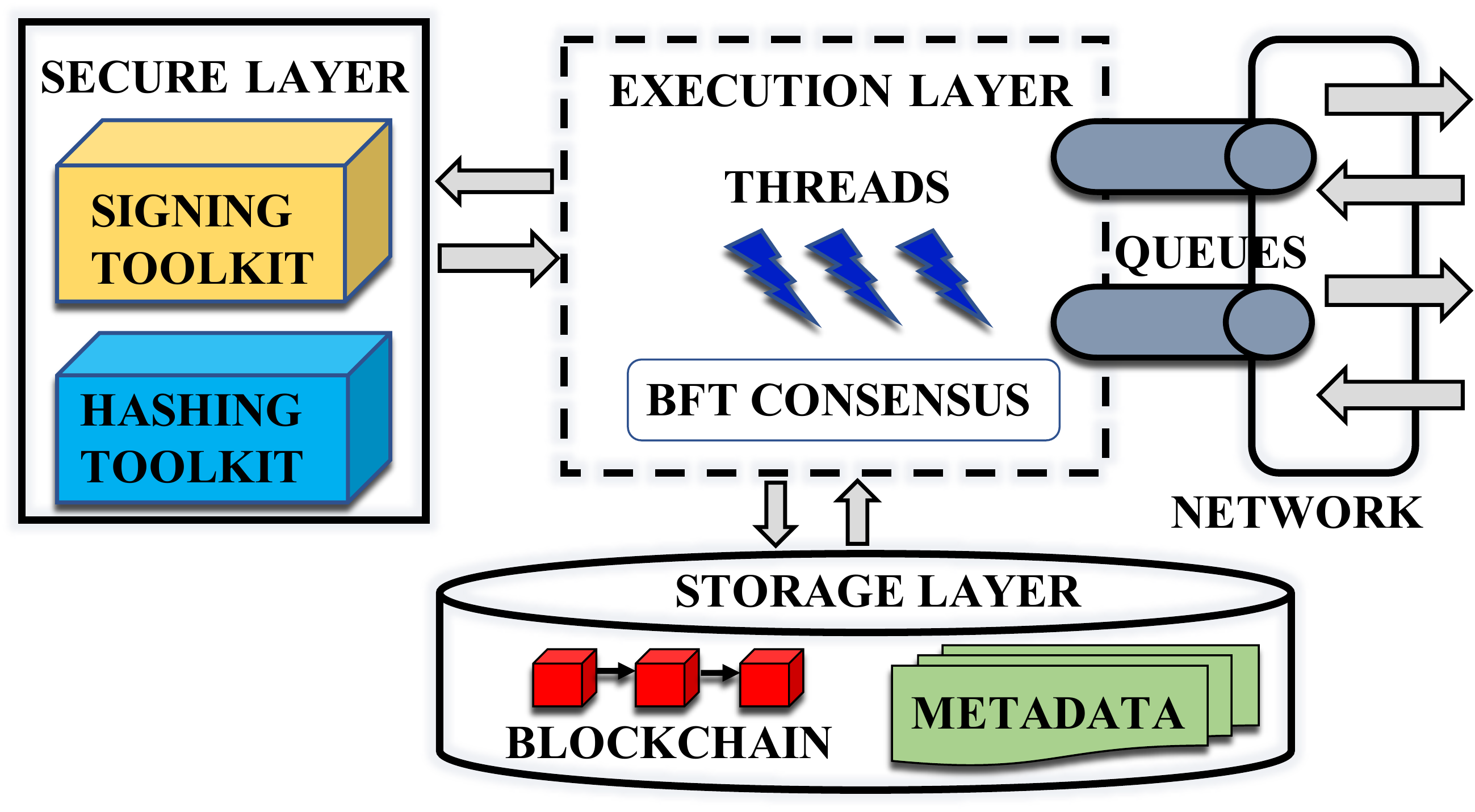}
\caption{\small Architecture of \expodb{}.}
\label{fig:expodb}
\end{figure*}

{\bf ResilientDB}~\citep{resilientdb,resilientdb-demo} 
is a state-of-the-art permissioned blockchain fabric, which is designed with the aim of 
fostering academic and industry research. 
\expodb{} also acts as a reliable test-bed to implement and evaluate  
enterprise-grade blockchain applications%
\footnote{%
\resilientdb{} is open-sourced at https://resilientdb.com and code is available at https://github.com/resilientdb.}.
\expodb{} evolved from the ExpoDB 
platform~\citep{expodb,easyc} which 
is an experimental research
platform to design and test 
emerging database technologies, agreement and concurrency 
control protocols. 

In Figure~\ref{fig:expodb}, we illustrate the overall architecture of \expodb{}, which
lays down an efficient client-server architecture. 
At the {\em application layer}, we allow multiple clients to co-exist, each of which creates 
its own requests. 
For this purpose, they can either employ an existing benchmark suite or design a 
{\em Smart Contract} suiting to the active application. 
Next, clients and replicas use the {\em transport layer} to exchange messages across the network.
\expodb{} also provides a {\em storage layer} where all the metadata corresponding to a request and 
the blockchain is stored.
At each replica, there is an {\em execution layer} where the underlying consensus protocol is run 
on the client request, and the request is ordered and executed.
During ordering, the {\em secure layer} provides support 
for cryptographic constructs.

\resilientdb{} is written entirely in C++ and provides a 
{\em graphical user interface} to ease user interaction with the system. 
Further, it also provide a {\em Dockerized} deployment that allows any user to 
experience and test the \resilientdb{} fabric (comprising of multiple replicas and clients) on its 
local machine.

\begin{figure}[t]
   \centering
   \includegraphics[width=\columnwidth,trim=0 3 0 21,clip]{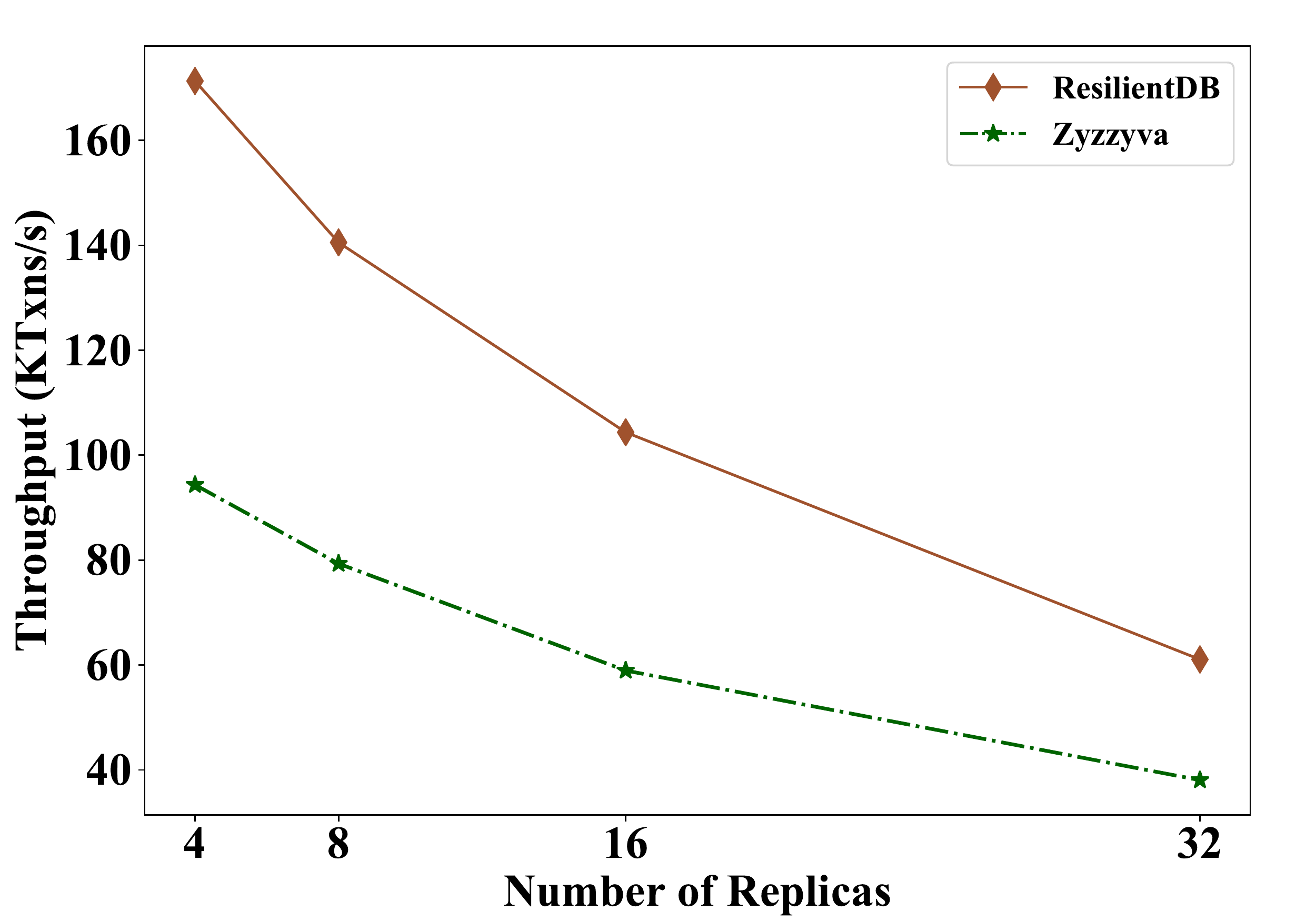}
   \caption{Two permissioned applications  employing distinct \bft{} protocols ($80$K clients per experiment).}
   \label{fig:compare-tput}
\end{figure}

The key motivation behind \resilientdb's design was to show that a {\em system-centric} 
permissioned blockchain fabric can outperform a {\em protocol-centric} blockchain fabric 
even if the former is made to employ a slow consensus protocol. 
To prove this claim, we refer to Figure~\ref{fig:compare-tput}, which compares the throughput achieved 
by two permissioned fabrics. 
In this figure, \resilientdb{} employs the slow \pbft{} protocol while the other fabric adopts the 
practices suggested in the paper BFTSmart~\citep{bftsmart} and employs the single-phase linear \zyzzyva{} protocol.
Despite this disadvantageous choice of consensus protocol, \resilientdb{} achieves a throughput of $175$K 
transactions per second, scales up to $32$ replicas, and attains up to $79\%$ 
more throughput.

\section{Future Directions for Research}
Although \bc{} technology is just a decade old, 
it gained majority of its momentum in the last five years. 
This allows us to render different elements of the \bc{} 
systems and achieve higher performance and throughput.
Some of the plausible directions to develop 
efficient \bc{} systems are: 
(i) reducing the communication messages, 
(ii) defining efficient block structure, 
(iii) improving the consensus algorithm, and 
(iv) designing secure light-weight cryptographic functions

Statistical and machine learning approaches have 
presented interesting solutions to automate key 
processes such as Face Recognition~\citep{facerecog},
Image classification~\citep{imageclassify}, 
Speech Recognition~\citep{speechrecog} and so on. 
The tools can be leveraged to facilitate easy and 
efficient consensus. 
The intuition behind this approach is to allow 
learning algorithms to select nodes, which are 
fit to act as a block creator and prune the 
rest from the list of possible creators.
The key observation behind such a design is that 
the nodes selected by the algorithm are predicted to 
be non-malicious.
Machine learning techniques can play an important role
in eliminating the human bias and inexperience.
To learn which nodes can act as block creators,  
a feature set, representative of the nodes, needs to 
be defined.
Some interesting features can be: geographical distance, 
cost of communication, 
available computational resources, 
available memory storage and so on.
These features would help in generating the 
dataset that would help to train and test the underlying 
machine learning model.
This model would be ran against new nodes that 
wish to join the associated \bc{} application. 

The programming languages and software engineering 
communities have developed several works that 
provide semantic guarantees to a language 
or an application~\citep{verdi, formalcompiler, 
cakeml}.
These works have tried to formally 
verify~\citep{formalverification, formalcompiler} 
the system using the principles of programming 
languages and techniques such as finite state 
automata, temporal logic and model 
checking~\citep{modelchecking, principlesmodelcheck}.
We believe similar analysis can be performed in the 
context of \bc{} applications.
Theorem provers (such as Z3~\citep{z3}) and proof 
assistants (such as COQ~\citep{coq}) could prove 
useful to define a certified \bc{} application.
A certified \bc{} application can help in 
stating theoretical bounds on the resources 
required to generate a block. 
Similarly, some of the \bc{} consensus has been 
shown to suffer from Denial of Service 
attacks~\citep{research-bitcoin}, and 
a formally verified \bc{} application can help  
realize such guarantees, if the underlying application 
provides such a claim.

\bibliographystyle{spbasic}  
\bibliography{benchbib} 

\end{document}